# LSMTCR: A Scalable Multi-Architecture Model for Epitope-Specific T Cell Receptor de novo Design

Ruihao Zhang[1], Xiao Liu[*]

*Tsinghua Shenzhen International Graduate School*

*Tsinghua University*

Shenzhen, China

* Corresponding authors

***Abstract*—Designing full-length, epitope-specific TCR αβ remains challenging due to vast sequence space, data biases and incomplete modeling of immunogenetic constraints. We present LSMTCR, a scalable multi-architecture framework that separates specificity from constraint learning to enable de novo, epitope-conditioned generation of paired, full-length TCRs. A diffusion BERT encoder learns time-conditioned epitope representations; enhanced GPT decoders, pretrained on CDR3β and transferred to CDR3α, generate chain-specific CDR3s under cross-modal conditioning with temperature-controlled diversity; and a gene-aware Transformer assembles complete α/β sequences by predicting V/J usage to ensure immunogenetic fidelity. Across TEP, MIRA, McPAS and our curated dataset, LSMTCR achieves higher predicted binding than baselines on most datasets, more faithfully recovers positional and length grammars, and delivers superior, temperature-tunable diversity. For α-chain generation, transfer learning improves predicted binding, length realism and diversity over representative methods. Full-length assembly from known or de novo CDR3s preserves k-mer spectra, yields low edit distances to references, and, in paired α/β co-modelling with epitope, attains higher pTM/ipTM than single-chain settings. LSMTCR outputs diverse, gene-contextualized, full-length TCR designs from epitope input alone, enabling high-throughput screening and iterative optimization.**

## I. INTRODUCTION

T cells survey the body for signs of infection and malignancy by exquisitely recognizing peptide antigens presented by major histocompatibility complex (MHC) molecules through their surface T cell receptors (TCRs) [1-3]. This molecular discrimination between self and non-self is executed by the paired α and β chains of the TCR, whose complementarity-determining regions (CDRs)—and in particular the hypervariable CDR3 loops—govern binding affinity and specificity to peptide–MHC (pMHC) complexes [4-8]. Engineered TCRs have shown clinical promise, with adoptive TCR-T therapies achieving notable efficacy in select hematological malignancies [9,10]. Yet, translating these successes to solid tumours and personalized immunotherapy remains challenging [11]. Tumour antigens in solid cancers are often heterogeneous and weakly immunogenic, frequently derived from self, increasing the likelihood of negative selection against useful receptors and amplifying the risks of off-target or cross-reactive toxicity [12-14]. Meanwhile, the combinatorial diversity of the natural TCR repertoire spans a staggering 10^15–10^61 possibilities, whereas wet-lab discovery remains low-throughput, costly and time-consuming [15-17]. As a result, the long-standing ambition of "epitope-to-full-length TCR αβ" de novo design has largely remained aspirational.

Recent advances in deep learning have opened new avenues for TCR engineering. With large-scale pretraining and conditional generation now converging with discriminative predictors, the field is shifting from purely predictive binding models to epitope-constrained sequence generation [18,19]. Representative approaches, such as GRATCR [20], couple an epitope encoder with a GPT-like generator via a post hoc "grafting" strategy, strengthening representation of CDR3β–epitope interactions but still yielding limited sequence diversity. Diffusion-based methods (for example, TCR-epiDiff [21]) improve reconstruction of known binding distributions and enhance diversity, yet often struggle with controlling length distributions and ensuring functional plausibility. Collectively, these efforts highlight the promise of generative frameworks for immune receptor design, while exposing key methodological and evaluation gaps: benchmarks frequently rely on fixed discriminators or proxy metrics that do not jointly capture specificity and diversity or align closely with biophysical reality [18,22]; training data are biased towards strong responders or easily observed events, constraining scale, coverage and out-of-distribution robustness [19,20]; and generation typically concentrates on CDR3β fragments, with limited explicit modeling of the α chain, V/J gene context, and full-length constraints critical to specificity, expression and manufacturability [20,21,23].

Achieving epitope-conditioned, full-length TCR αβ design therefore requires a step-change in both scale and modeling strategy. An effective framework must learn epitope–receptor interaction patterns while respecting immunogenetic constraints arising from V(D)J recombination, chain pairing and repertoire statistics, and should directly output paired, full-length sequences suitable for experimental construction [24-27]. Methodologically, such a system should combine the bidirectional context modeling strengths of masked encoders with the controllability and sampling efficiency of autoregressive decoders; leverage data-efficient pretraining while maintaining generative flexibility; and support rigorous, multidimensional evaluation grounded in biological plausibility.

Here we present LSMTCR, a scalable, multi-architecture model for epitope-specific de novo design of full-length TCR αβ. LSMTCR separates target and constraint modeling—learning epitope-conditioned CDR3 specificity while enforcing V/J choice, chain pairing and immunogenetic consistency—and then fuses them through staged training and conditioning. First, a diffusion-enhanced BERT encoder learns robust, time-conditioned epitope representations, while GPT-based decoders are pretrained on CDR3β and adapted to CDR3α via transfer learning. Second, a cross-modal conditioning mechanism aligns epitopes and CDR3s in a shared embedding space, reinforced by time conditioning, length perturbations and a noise curriculum to improve robustness to weak supervision and unseen epitopes. Third, controlled decoding with temperature scheduling yields diverse and tunable CDR3α/β candidates given a specified epitope. Finally, a gene-aware Transformer assembler predicts matching V/J genes from (epitope, CDR3α, CDR3β) and synthesizes full-length TCRA/TCRB, ensuring immunogenetic fidelity and engineering feasibility.

Compared to existing approaches, LSMTCR advances the state of the art along several axes. It integrates diffusion-style epitope encoding with conditional autoregressive generation, enhancing robustness without sacrificing controllability; aligns epitopes and CDR3s in a shared space to improve generalization beyond training distributions; and introduces retrieval-augmented, soft-constrained decoding at inference to mitigate dataset biases and improve out-of-distribution performance. Crucially, LSMTCR completes the pipeline from fragments to full-length receptors by first generating diverse CDR3α/β and then jointly predicting V/J usage and assembling complete α/β chains consistent with V/J statistics and chain pairing distributions. This produces candidates that better balance specificity, expression feasibility and practical constructability. By accepting only an input epitope, LSMTCR can rapidly propose diverse, gene-contextualized, full-length TCR αβ designs, enabling high-throughput in vitro screening, iterative optimization and mechanistic interrogation—bringing epitope-to-receptor design closer to routine application in precision immunotherapy.

# II. RESULTS

## LSMTCR overview

We introduce LSMTCR, a scalable, multi-architecture framework for epitope-specific, end-to-end de novo generation of full-length TCR α and β chains. The system integrates three complementary modules—Epitope-BERT, CDR3-GPT, and TCR-Transformer—tailored respectively to capture epitope features, model CDR3 sequence distributions, and learn the hierarchical dependencies that govern full-length TCRs.

In pretraining, Epitope-BERT learns contextual representations of epitopes from large-scale datasets [20], while CDR3-GPT models the distribution of CDR3β sequences and incorporates a masked-reconstruction objective to improve generalization. Given the relative scarcity of CDR3α data, we adopt a transfer learning strategy: we first pretrain on abundant CDR3β corpora and then adapt the model to CDR3α, optimizing representations for α-chain specificity. During fine-tuning, we condition on epitope–CDR3β and epitope–CDR3α interaction data to capture chain-specific binding patterns. We then introduce a conditional GPT decoder—partially freezing pretrained weights—to generate CDR3 sequences under explicit epitope conditioning (Fig. 1).

To produce full-length receptors, we further employ a TCR-Transformer that learns the layered dependencies among CDR3 segments, V/J gene usage, and complete α/β sequences through large-scale pretraining, using data drawn from our curated dataset. At inference, the model first simultaneously predicts V and J genes consistent with the generated CDR3s and then assembles full-length TCRα and TCRβ chains in a stepwise manner. This staged, modular design enables LSMTCR to generate biologically coherent, sequence-level accurate full-length TCRs with high efficiency, providing a practical and extensible platform for epitope-targeted TCR design and immunotherapeutic development.

## Comparison with existing CDR3β generative models in binding probability

To assess generative reliability, we benchmarked LSMTCR against deep learning models on the CDR3 generation task using a unified evaluation pipeline. Due to the lower sequence diversity of GRATCR model, we generated 20 TCRs per epitope and selected 10 non-redundant CDR3s for evaluation. For each model's outputs, we estimated epitope–CDR3 binding probability with NetTCR [28], a widely used deep learning model for predicting CDR3–epitope interactions, and conducted comparisons across multiple datasets (TEP [29], MIRA [30], McPAS [31], and our curated cohort). We also used another two deep learning models with different architectures, ATM-TCR [32] and TEPCAM [33], to predict the binding probability between our model–generated CDR3β sequences and epitopes across multiple datasets. Negative examples were constructed by shuffling training pairs to ensure consistent calibration of the discriminator.

Across TEP, MIRA, and our dataset, LSMTCR achieved the highest predicted binding probabilities as indicated by NETTCR (Fig. 2a). On McPAS, performance was slightly below that of GRATCR; however, LSMTCR's generated CDR3β sequences exhibited a more concentrated distribution of high binding scores with fewer low-affinity outliers, as evident from the box-plot distributions (Fig. 2b). TCR-epidiff produced poor CDR3–epitope binding because its generated CDR3s deviated from the fine-tuning data (Fig. 2c), whereas our model effectively captured the CDR3 characteristics present in the fine-tuning set. Across all datasets, the ATM-TCR model's average prediction for LSMTCR was 100%, and the TEPCAM model's average prediction for LSMTCR was 99.99%. These results indicate that LSMTCR not only matches or surpasses existing approaches on most benchmarks, but also produces outputs with tighter binding quality control—an attribute desirable for downstream screening and experimental validation.

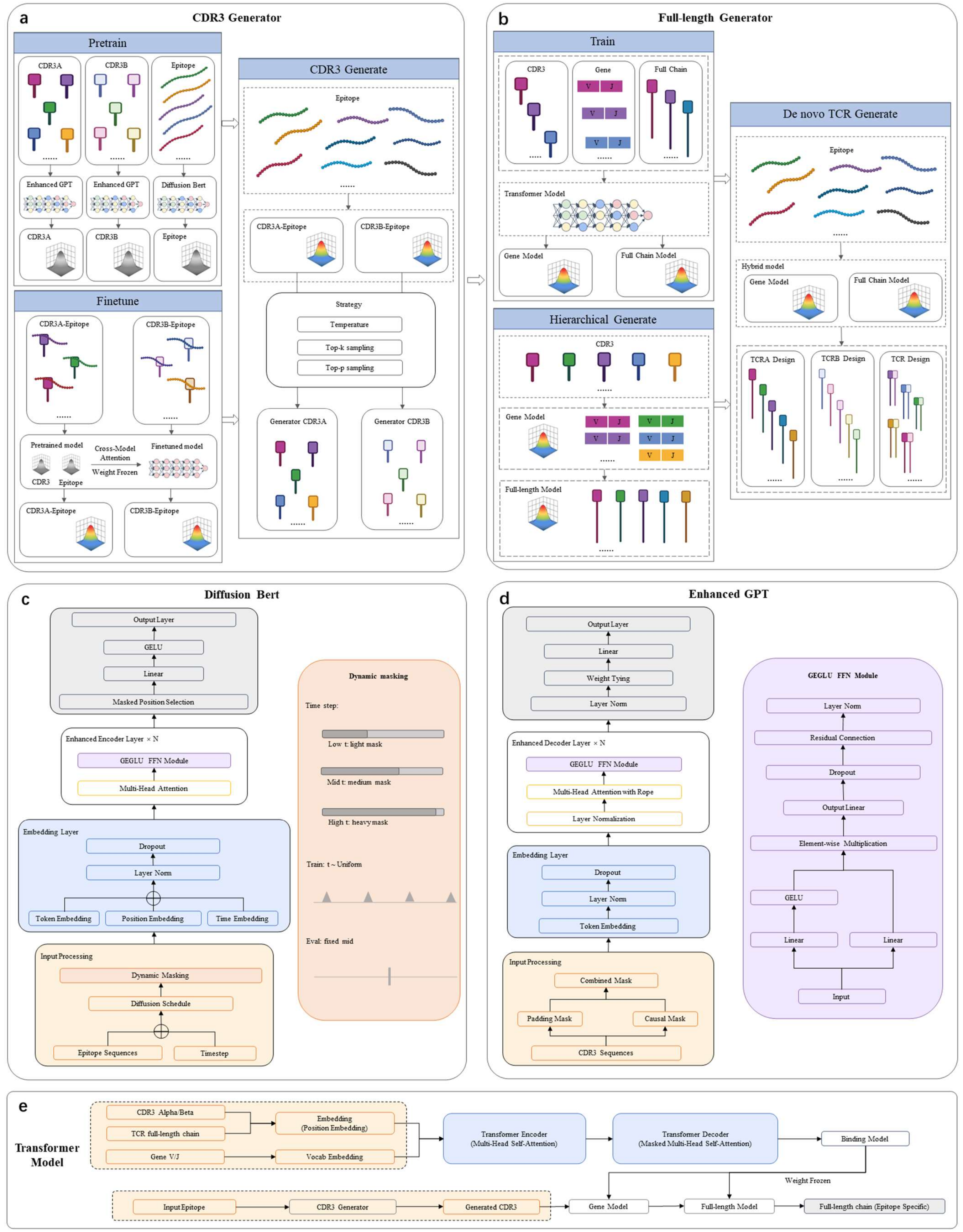

**Fig. 1 | Illustration of the LSMTCR Model.** a, CDR3 generator: pre-training (diffusion-enhanced BERT; enhanced GPT) and cross-model fine-tuning for epitope–CDR3 binding; conditional CDR3$\alpha/\beta$ generation. b, Full-length generator: trained on V/J genes, CDR3 and full-length sequences; hierarchical generation (V/J from CDR3, then full-length); end-to-end TCR design. c, Diffusion-enhanced BERT for epitope encoding with dynamic masking. d, Enhanced GPT for CDR3 with multi-head attention and GEGLU FFN module. e, TCR-generation transformer conditioned on CDR3, full-length context and V/J metadata.

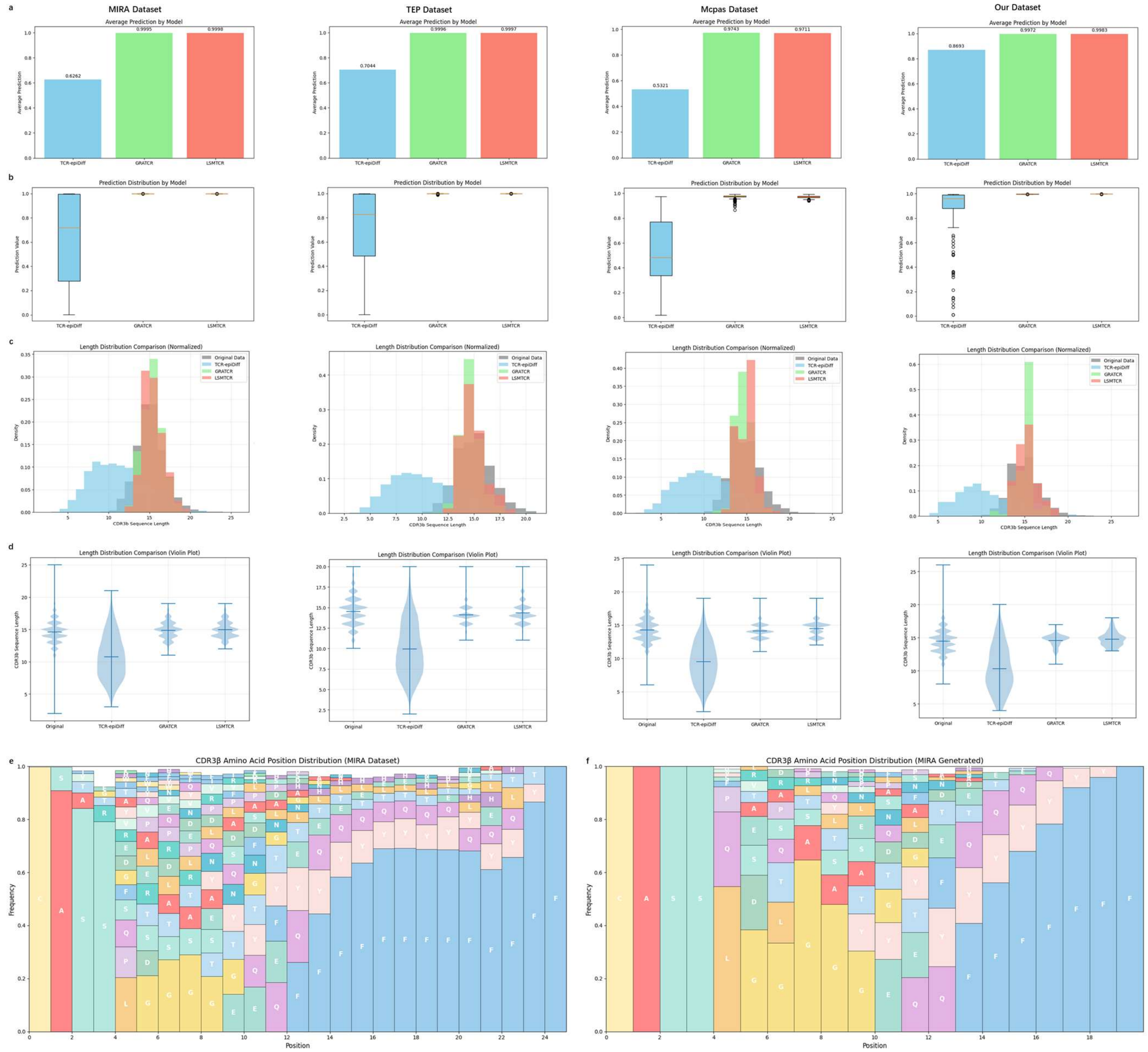

**Fig. 2 | The performance of LSMTCR and other existing tools.** a, NetTCR-predicted epitope–CDR3β binding probabilities for sequences generated by LSMTCR, TCR-epidiff and GRATR across datasets. b, Distributions of predicted binding probabilities across models and datasets. c, Histograms of generated CDR3β lengths versus empirical lengths, assessing concordance with the original data. d, Violin plots of generated versus empirical CDR3β lengths. e, Position-wise amino acid frequencies for LSMTCR-generated CDR3β on MIRA compared with the original data.

## Comparison of generated CDR3β amino acid distributions with background

To determine whether LSMTCR captures intrinsic properties of CDR3β sequences, we profiled the position-specific amino acid distributions in the generated repertoires and compared them to the corresponding distributions measured from the background dataset, that is, the position-specific frequencies observed in real background data. Given its large sample size over 43000 and its relative independence, we focus on the MIRA cohort, contrasting native CDR3β positional amino-acid frequencies with those derived from model-generated sequences.

While the generative distributions show reduced amino acid variety at certain positions—consistent with sampling constraints and model regularization—the global positional profiles closely track the empirical background (Fig. 2e). This convergence indicates that LSMTCR effectively learns the positional amino acid preferences and salient motifs of CDR3β, preserving sequence grammar at the level relevant for structural and functional interpretation.

## Comparison with existing generative models in CDR3β sequence length

Alignment of generated and empirical length distributions is a key indicator of whether a model faithfully captures the constraints governing CDR3–epitope recognition. We therefore compared the length profiles of generated CDR3β sequences with those of the originating datasets across multiple cohorts, noting that natural CDR3β lengths typically occupy a constrained range (approximately 8-15 amino acids [34]) rather than spanning arbitrary values.

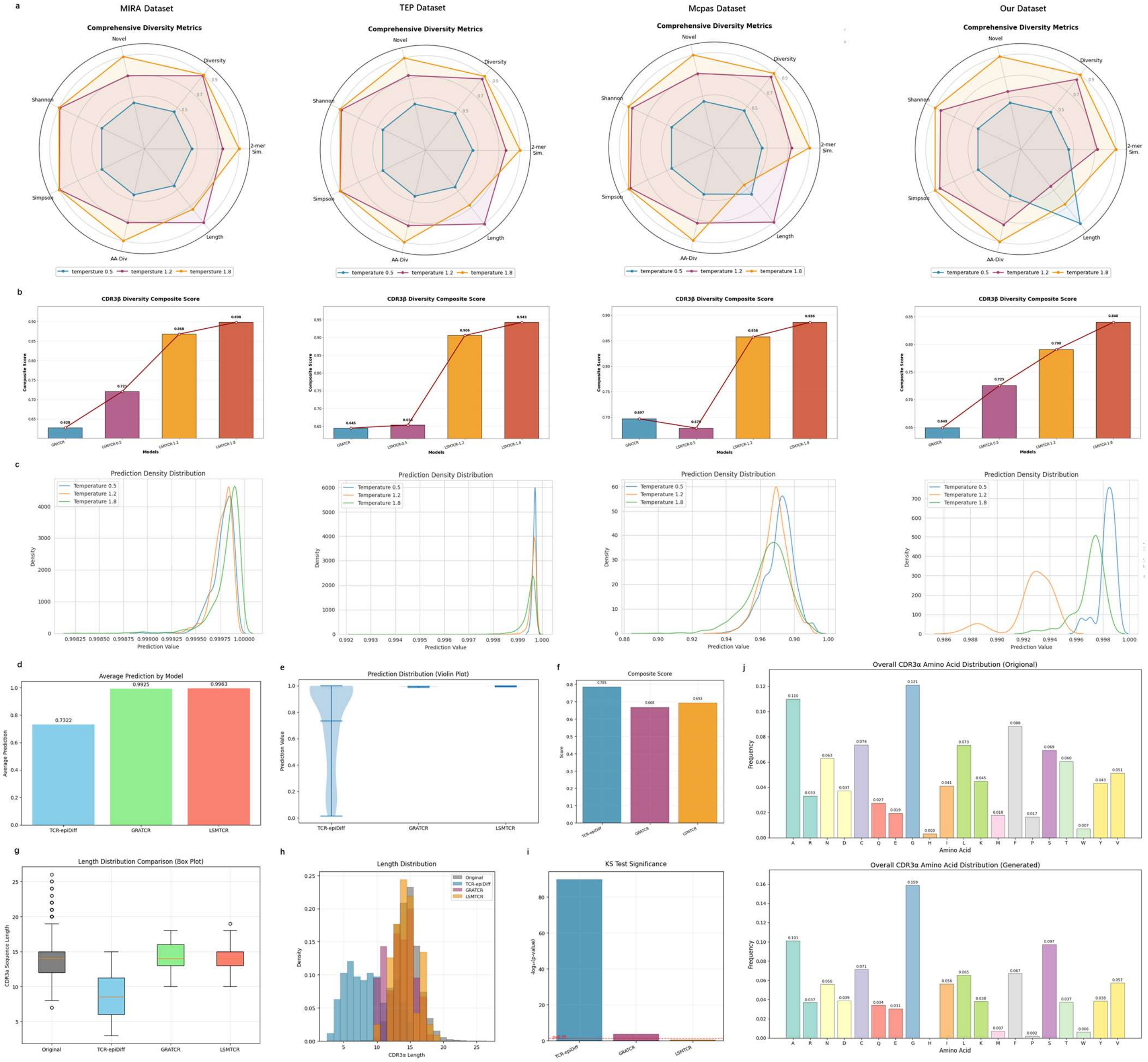

**Fig. 3 | Temperature-tuned diversity in LSMTCR and benchmarking of CDR3α models.** a, LSMTCR performance across datasets under varying temperatures, evaluated by 2-mer Jaccard, diversity ratio, novel ratio, Shannon and Simpson indices, AA-Div and length realism score. b, CDR3β Diversity Composite Score of LSMTCR across datasets (weighted from the seven metrics) compared with GRATCR. c, Binding probability distributions of LSMTCR-generated CDR3β to epitopes under different temperatures. d, Epitope–CDR3α binding probabilities generated by different models after training on our curated CDR3α dataset. e, Distributions of epitope–CDR3α binding probabilities for sequences generated by different models on our curated CDR3α dataset. f, CDR3α Diversity Composite Score for sequences from different models on our curated CDR3α dataset. g,h, Length distributions of generated CDR3α on our curated CDR3α dataset shown as boxplots and bar charts. i, KS test significance for generated CDR3α from different models. j, Amino acid frequency distributions in the original CDR3α data versus those generated by LSMTCR.

LSMTCR closely recapitulates the empirical length distributions, whereas TCR-epiDiff tends to produce shorter-than-expected sequences, reducing practical utility (Fig. 2c). Violin plots further show that LSMTCR achieves the best fit to the native distribution, exhibiting broad support across the observed range, while GRATCR concentrates disproportionately on a few lengths (Fig. 2d). Taken together, these results indicate that LSMTCR most accurately reproduces the natural length grammar of CDR3β, reinforcing its ability to learn biologically grounded sequence constraints relevant to binding and downstream usability.

## Temperature-controlled diversity in CDR3β generation

Sampling temperature is a primary control for modulating diversity in LSMTCR, with direct relevance to discovery and translational applications. By tuning temperature value, one can enrich CDR3β repositories for downstream binding prediction while assembling broader candidate pools for experimental screening. We systematically varied temperature across multiple datasets and quantified the fidelity–diversity trade-off using complementary metrics: 2-mer Jaccard (concordance of local dipeptide motifs between generated and empirical repertoires), diversity ratio (uniqueness after deduplication), novel ratio (proportion absent from the reference, capturing exploratory reach), Shannon and Simpson indices (distributional evenness and attenuation of dominant clones, reported relative to empirical baselines), AA-Div (amino-acid compositional richness and balance), and a length realism score (deviation of mean length from the native distribution).

As temperature increased in TEP, McPAS, MIRA and our curated cohort, generated repertoires showed broader coverage of local motifs and a more balanced amino-acid composition, together with higher uniqueness and a larger share of genuinely novel sequences. Entropy-based evenness likewise rose, indicating more uniform frequency spectra and reduced dominance of a few motifs (Fig. 3a). Length realism was more sensitive and often non-monotonic: with increasing temperature, the mean length either rose and then fell or decreased steadily, reflecting weakened control over the natural CDR3β length grammar at high temperatures.

Predicted binding probability distributions clarify the attendant trade-offs. Lower temperatures concentrate samples near high-affinity modes—yielding higher binding probabilities but reduced diversity—whereas higher temperatures promote exploration across sequence space, increasing novelty and evenness at a modest cost to predicted affinity (Fig. 3c). These observations establish temperature as a practical lever for task-tailored generation, favouring fidelity and prioritization at low settings, and exploration and library expansion at higher ones.

## Comparison with existing generative models in CDR3β sequence diversity

Sequence diversity is a critical dimension of generative performance with direct implications for clinical discovery. We compared LSMTCR with GRATCR by generating CDR3β repertoires across multiple datasets under varying temperature settings and quantifying diversity using a composite score that integrates multiple radar-plot metrics. While the McPAS cohort did not show a monotonic increase, most datasets exhibited a clear trend: higher temperatures yielded more diverse repertoires for LSMTCR, reflecting broader exploration of sequence space without collapsing onto a few high-frequency modes (Fig. 3b). Across matched temperature regimes, LSMTCR consistently achieved higher Diversity Composite Scores than GRATCR, indicating that our framework not only maintains strong predicted binding but also delivers superior diversity—an advantage for constructing robust candidate libraries and mitigating overfitting to dataset-specific patterns.

## Evaluation of CDR3α generation

Leveraging the α-chain data in our curated cohort, we conducted a dedicated evaluation of CDR3α generation. Predicted epitope–TCR binding analyses show that LSMTCR outperforms GRATCR and TCR-epiDiff on CDR3α task, underscoring the benefits of our transfer-learning pretraining and the tailored decoder architecture (Fig. 3d, e). Length profiles of generated CDR3α sequences closely match the empirical distribution, yielding candidates in the most usable range, and diversity exceeds that of GRATCR across matched settings—indicating that LSMTCR maintains binding quality without sacrificing repertoire breadth. Position-specific amino-acid frequencies in generated sequences follow the global trends of the reference data, consistent with faithful recovery of the compositional and motif grammar characteristic of CDR3α (Fig. 3j). The KS test Significance of LSMTCR demonstrates that LSMTCR achieves a favourable balance between affinity, realism and diversity for α-chain design.

## Evaluation of full-length TCR generation from known CDR3

The core of LSMTCR's TCR generator is a Transformer model. We first train it on our curated dataset containing complete information—CDR3, full-length sequences, and V/J gene annotations—and then adopt a staged generation strategy: the model first produces the genes corresponding to the CDR3 region, and subsequently generates the full-length TCR sequence conditioned on these genes. To assess the quality of generated full-length sequences with Transformer model, using our curated dataset with 18 distinct epitopes, we also trained a Transformer-based classifier of epitope–TCR binding and validated its reliability by constructing negatives via shuffled training pairs and evaluating on the all-positive test set. The Transformer-based classifier achieved ACC, Recall, F1 and Specificity of 1.0, supporting its use as a calibration tool. We then applied it to long-chain sequences assembled from known CDR3s: both TCRα and TCRβ exhibited predicted binding probabilities exceeding 99.99%, indicating that full-length sequences derived from background CDR3s retain strong epitope compatibility.

We next compared length distributions between generated and reference full-length chains, observing close agreement and thereby supporting length realism (Fig. 4a,b). Multiple sequence similarity metrics further corroborated fidelity. Exact-match rates exceeded 0.6 for TCRα and 0.4 for TCRβ; normalized Hamming distances were below 0.1 and 0.2, respectively; and Levenshtein distances remained below 0.2 (α) and 0.3 (β), consistent with limited edit operations and alignment to empirical length statistics. Local motif concordance, quantified by 3-mer set Jaccard similarity, was high—above 0.7 for α and 0.5 for β—indicating preservation of short-range "micro-grammar" even when global edits were present (Fig. 4c).

Frequency-concordance analyses reinforced these findings. For both α and β chains, Jensen–Shannon divergence between generated and reference k-mer spectra (k=2, 3) was small, with point clouds tightly distributed along the $y = x$ diagonal, demonstrating recovery of empirical frequency profiles without over-amplifying rare fragments. Consistently low 3-mer frequency differences further indicated strong vocabulary and spectral agreement. Together, these results show that, when seeded with background CDR3s, the TCR-Transformer assembles full-length α/β chains that are distributionally faithful, motif-consistent and predicted to bind their target epitopes with high confidence (Fig. 4e,f).

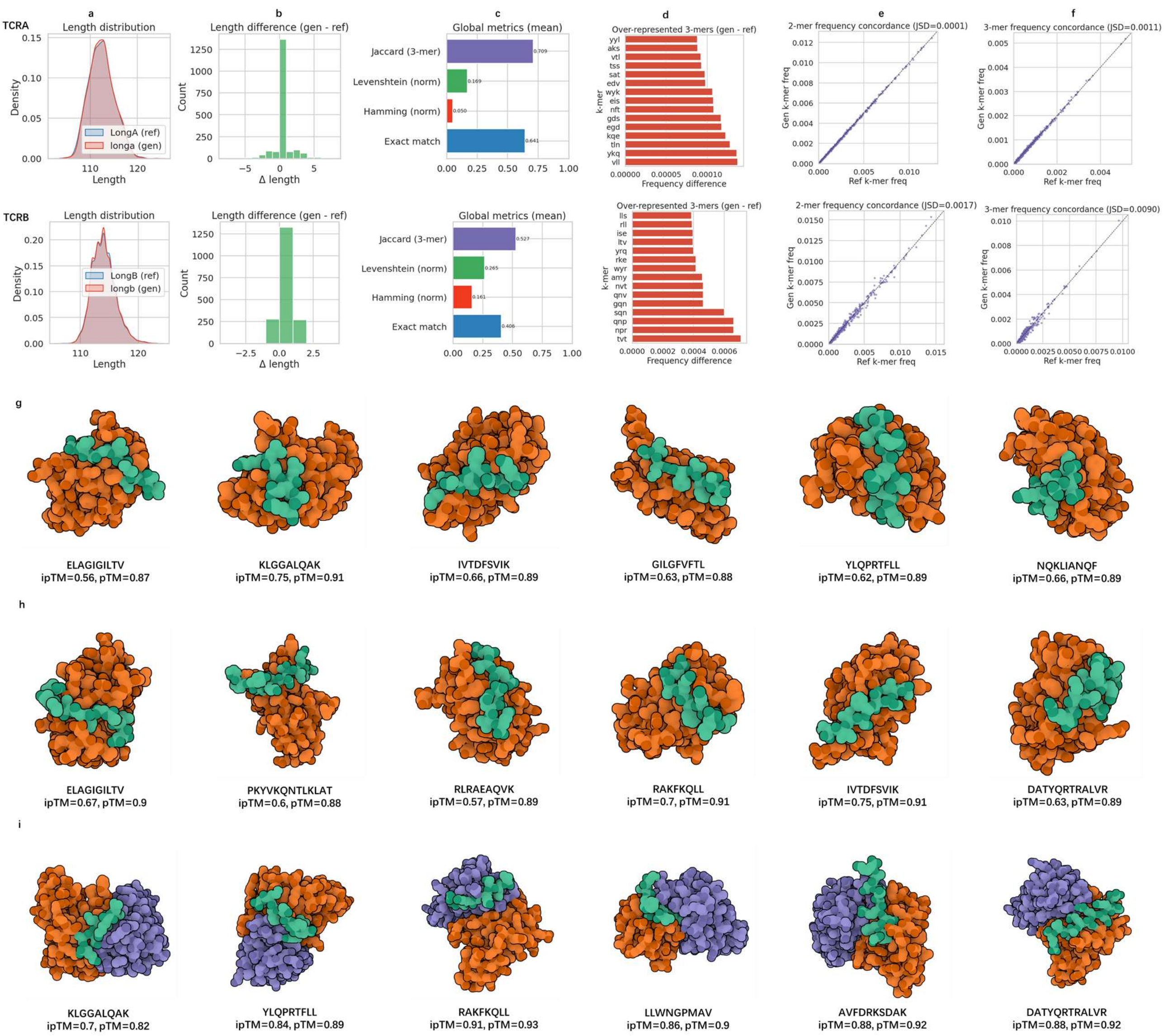

**Fig. 4 | Evaluation of full-length TCR generation from background CDR3 and structural assessment.** a, Length distributions of generated TCRα/β full-length sequences from known CDR3s. b, Length differences between generated and background TCRα/β sequences. c, Similarity to background: 3-mer set Jaccard, Levenshtein and Hamming distances, and exact-match rate. d, 3-mer vocabulary divergence between generated and background sequences. e,f, 2-mer and 3-mer frequency concordance with background. g,h, AlphaFold structural visualizations of epitope binding for generated TCRα and TCRβ full-length chains. i, AlphaFold structure of epitope binding for paired generated TCRα/β chains.

## Structural assessment of generated TCR from known CDR3

Following full-length sequence design, we assessed binding configurations using AlphaFold-based complex modeling [35]. Starting from known CDR3s, we assembled full-length chains and predicted structures under two settings: single-chain complexation with the epitope and paired α/β co-modeling with the same epitope. The paired models consistently achieved higher pTM and ipTM scores than their single-chain counterparts (Fig. 4g,h). In particular, pTM values exceeding 0.5 suggest globally plausible folds, and ipTM values above 0.8 indicate reliable subunit orientations and interface organization [36,37]. In our analyses, the α/β co-models more frequently reached or approached these thresholds, pointing to greater confidence in both overall topology and binding interface.

This behaviour accords with the biophysics of TCR recognition: specificity and affinity emerge from the composite interface formed by the α/β heterodimer. Single-chain modeling with pMHC lacks the geometric and electrostatic constraints imposed by the partner chain, leading to reduced ipTM and greater drift of the interface. By contrast, co-modeling supplies a more complete complementary surface and stereochemical context, stabilizing interface packing and ligand pose, and thereby improving both ipTM and pTM. While single-chain predictions can serve as a rapid screening aid, the paired setting markedly increases interface determinacy and global fold credibility, indicating that our generated full-length sequences are more likely to adopt stable, biologically plausible binding modes when modeled as α/β heterodimers.

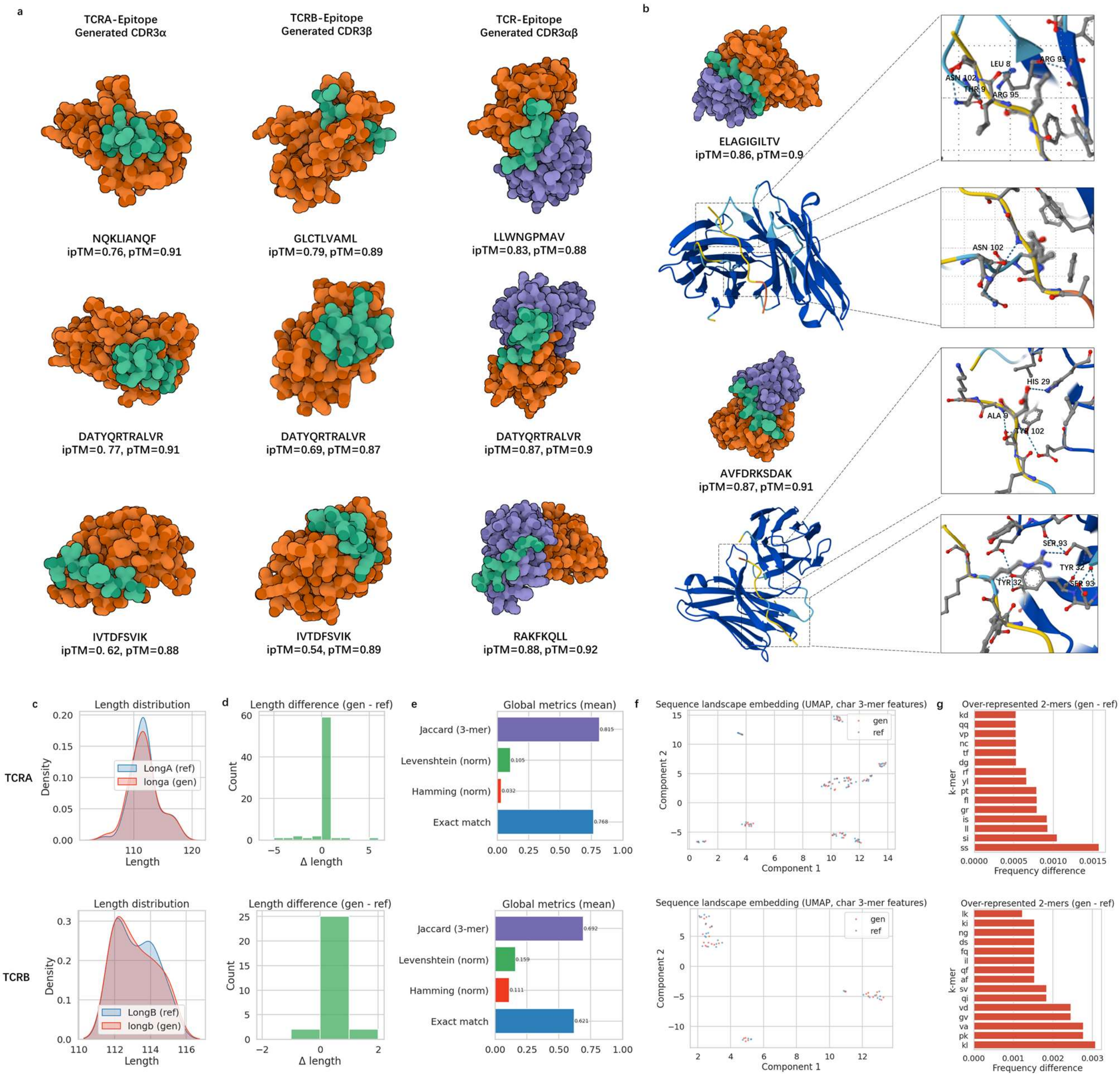

**Fig. 5 | Evaluation of full-length TCR generation from de novo generated CDR3 and structural assessment.** a, AlphaFold structural visualizations of epitope binding for LSMTCR-designed de novo TCRα, TCRβ and paired αβ chains. b, Visualization of interfacial contacts/bonds at the epitope–TCRαβ binding site. c, Length distribution curves of de novo TCRα/β versus background references. d, Length differences between de novo and background TCRα/β sequences. e, Similarity to background: 3-mer set Jaccard, Levenshtein and Hamming distances, and exact-match rate. f, UMAP embedding of de novo and background TCRα/β sequences. g, 2-mer vocabulary divergence between de novo and background sequences.

## Evaluation of full-length TCR generation from de novo CDR3

We next assessed de novo performance by first generating CDR3α and CDR3β sequences directly from epitope conditioning and then assembling full-length TCRα and TCRβ with the TCR-Transformer. The Transformer-based classifier indicated that both chains achieved predicted binding probabilities above 99.99%, supporting the functional plausibility of the end-to-end designs.

Length distributions of the de novo full-length sequences closely matched those of empirical references, indicating realistic chain assembly. Multiple similarity metrics corroborated sequence-level fidelity: exact-match rates exceeded 0.7 for α and 0.6 for β; normalized Hamming distances remained below 0.1 (α) and 0.2 (β); and Levenshtein distances were below 0.2 across both chains, consistent with minimal edit operations relative to reference (Fig. 5e). Local motif preservation was strong, with 3-mer set Jaccard similarities above 0.8 for α and 0.6 for β, indicating high concordance of short-range "micro-grammar" even when minor global edits were present.

Distributional analyses reinforced these findings. UMAP embeddings of α and β repertoires showed substantial overlap between de novo and reference sequences, indicating that epitope-conditioned designs occupy the same manifold as natural chains (Fig. 5f). Concordance of k-mer spectra (k=2) was high, with very small frequency differences relative to background, further evidencing strong vocabulary and spectral agreement. Together, these results demonstrate that LSMTCR can generate full-length α/β receptors from scratch that are length-realistic, compositionally faithful and predicted to bind their target epitopes with high confidence.

## Structural assessment of generated TCR from de novo CDR3

After assembling full-length receptors from epitope-conditioned, de novo CDR3s, we evaluated binding configurations using AlphaFold-based complex modeling under single-chain (α or β with epitope) and paired α/β co-modeling settings. The paired models consistently outperformed single-chain setups, achieving higher pTM and ipTM scores and more frequently meeting indicative thresholds (pTM > 0.5 for globally plausible folds; ipTM > 0.8 for reliable subunit orientation and interface organization). These gains reflect the biophysical reality that TCR specificity and affinity emerge from the composite α/β interface: co-modeling supplies a complete complementary surface and stereochemical context, stabilizing interface packing and ligand pose (Fig. 5a).

To further interrogate interface credibility, we examined two representative α/β complexes generated from de novo CDR3s. Both exhibited coherent multi-site contacts between the TCR and the epitope across the binding cleft, consistent with canonical docking geometries and supporting the practical usability of the designed receptors (Fig. 5b). Collectively, the structural analyses indicate that end-to-end designs produced by LSMTCR are predisposed to adopt stable, biologically plausible binding modes when modeled as heterodimeric complexes.

# III. Material and Mehods

## Datasets

We collected publicly available resources spanning TEP [29], MIRA [30] and McPAS [31], and assembled an additional curated cohort comprising 20,200 epitope–TCR pairs with complete chain and gene annotations from VDJdb [38] and 10xgenomics [39]. For pretraining the CDR3 generators, we leveraged large-scale corpora of approximately 1.5 million epitopes and 3 million TCR sequences drawn from public repositories [20]. For each dataset, we performed fine-tuning and generation independently. Specifically, each dataset was split 80:20 for training versus evaluation (or training versus generation, as appropriate), and all fine-tuning and generation were conducted within the same dataset to prevent cross-dataset leakage and ensure reproducibility. CDR3β generation was benchmarked across all four cohorts, whereas CDR3α and full-length chain generation were evaluated on the curated dataset, which provides matched α/β and V/J gene context required for end-to-end assembly.

## Diffusion-Enhanced BERT for Epitope Representation Learning

We developed a time-conditioned BERT encoder that performs masked language modeling under a progressive difficulty curriculum inspired by diffusion processes. Rather than training on a fixed masking ratio, our approach conditions the model on an explicit timestep variable $t$ that controls corruption severity, transforming standard MLM from a single-difficulty denoising task into a multi-stage reconstruction process across varying noise levels (Fig. 1c).

**Time-Conditioned Embeddings.** For a tokenized input sequence $\mathrm{x} = (x_1, \cdots x_s)$ with maximum length $S \leq \mathrm{S}_{\max}$ and a diffusion step $t \in \{1, \cdots, T\}$, token representations combine three embedding components: learned token embeddings, learned positional embeddings, and time-aware embeddings that encode the current corruption level:

$$E_{x,t} = E_{tok}(x) + E_{pos}(1{:}S) + E_{time}(t) \tag{1}$$

The time embedding $E_{time}(t)$ is implemented as either a learned lookup table of dimensions $(T+1) \times D$ or fixed sinusoidal positional encodings of equivalent shape, selectable via hyperparameter. This embedding is broadcast across all sequence positions, expanding from shape [B, D] to [B, S, D] to match the sequence dimension. Padding tokens (index 0) have their embeddings zeroed to eliminate spurious signals, and attention masks prevent the model from attending to padded positions. We found that injecting temporal information at the embedding stage effectively propagates corruption awareness throughout all encoder layers without requiring architectural modifications.

**Corruption Schedule and Adaptive Masking.** Training employs a strictly linear corruption schedule that monotonically increases masking difficulty over timesteps. At each optimization step, we sample $t$ uniformly from $\{0, \cdots, T\}$ and compute the masking proportion $p(t)$ as:

$$p(t) = P_{\min} + (P_{max} - P_{\min}) \cdot \frac{t}{T} \tag{2}$$

where $P_{\min}$ and $P_{max}$ define the minimum and maximum corruption levels, respectively. To maintain compatibility with our data preprocessing pipeline, which preselects $M$ candidate positions per sequence under a reference ratio $p_{ref} = 0.15$, we activate a subset of size:

$$m(t) = clamp(round(M \cdot p(t) / p_{ref}), 1, M) \tag{3}$$

This formulation ensures monotonic corruption increase while preserving alignment with preselected masking indices, avoiding computational overhead from position resampling. The parameter $P_{\min}$ establishes an "easy" regime comparable to conventional MLM ratios, while $P_{max}$ defines the most challenging reconstruction scenario; T controls curriculum granularity.

**Encoder architecture.** The model comprises $N$ identical Transformer encoder layers with hidden dimension $D = 768$ and $H = 12$ attention heads, each using key and value dimensions $d_k = d_v = 64$. Self-attention applies scaled dot-product mechanisms with padding masks derived from input sequences ($pad_{id} = 0$), followed by residual connections, dropout regularization, and layer normalization. The feed-forward component replaces standard two-layer MLPs with GEGLU [40] activation to enhance nonlinearity and optimization stability in domain-specific biological sequences:

$$Y = GELU(X \cdot W_a + b_a) \odot (X \cdot W_b + b_b) \tag{4}$$

$$FFN(X) = LayerNorm(Dropout(Y \cdot W_0 + b_o) + X) \tag{5}$$

where $W_a, W_b \in R^{\{D\times D_{ff}\}}$ and $W_o \in R^{\{D_{ff}\times D\}}$ with expansion ratio $D_{ff} = 4D$. The GEGLU gate provides multiplicative feature modulation, yielding improved token-wise selectivity without increasing model depth. All linear transformations use Xavier uniform initialization, with dropout applied to attention and feed-forward outputs (Fig. 1c).

**Masked decoding with tied input-output embeddings.** The model predicts exclusively at masked positions to focus learning on reconstruction tasks. Given encoder output $H \in R^{\{B\times S\times D\}}$ at timestep t and masked position set $M_t$, we extract masked hidden states, apply a GELU-activated projection, and decode using the transposed token embedding matrix (weight tying):

$$z_i = GELU(H(i)\cdot W_c + b_c) \quad (6)$$

$$\log it_i = z_i \cdot W_E^T, \qquad \forall i \in M_t \quad (7)$$

where $W_C \in R^{\{D\times D\}}$ provides a learned transformation and $W_E$ represents the shared input-output embedding matrix. Weight tying reduces parameter count while maintaining consistent lexical geometry between input and output representations.

**Training Objective and Optimization.** The training loss computes cross-entropy over masked targets at the sampled diffusion step $t$, averaged across masked positions:

$$L_{MLM}(t) = -(1/|M_t|)\sum_{i\in M_t} \log\, P_\theta\big(x_i|x_{setminus\ M_t,t}\big) \quad (8)$$

The overall objective marginalizes over timesteps through uniform sampling:

$$L = E_{t\sim \mathrm{Unif}\{1,\cdots T\}}[L_{MLM}(t)] \quad (9)$$

This formulation trains a single parameter set to operate robustly across the full spectrum of corruption levels, analogous to denoising diffusion adapted for discrete sequences but implemented within an encoder-only architecture.

We optimize parameters using AdamW with a linear learning rate schedule including 10% warmup steps. Mixed-precision training and distributed data parallelism are managed through the Accelerate library. Reproducibility is ensured by fixing random seeds across Python, NumPy, and PyTorch environments, disabling cuDNN benchmarking, and enabling deterministic operations. Validation employs the same reconstruction objective evaluated at mid-level corruption $t_{eval} = \lfloor T/2 \rfloor$ to provide consistent intermediate difficulty assessment across training epochs.

## Enhanced GPT Model for CDR3 Representation Learning

We developed an autoregressive decoder that adapts the GPT architecture to short, domain-specific biological sequences through three key innovations: pre-normalization Transformer blocks, rotary positional embeddings (RoPE) for geometry-aware attention, and gated GEGLU feed-forward networks for enhanced nonlinearity [40,41]. The model employs causal next-token prediction while incorporating padding-aware masking and weight-tied output projections to maintain coherent lexical geometry across input and output spaces.

**Input Representation and Attention Masking.** For a tokenized input sequence $x = (x_1, \cdots x_s)$ with padding identifier 0, we construct a composite attention mask that simultaneously enforces autoregressive causality and excludes padding positions. The causal component implements an upper-triangular mask ensuring that token $x_j$ depends only on preceding tokens $\{x_1, \cdots x_{j-1}\}$, while the padding component prevents attention to invalid positions:

$$M_{causal}(\mathrm{i}, j) = 1\ if\ i < j, \qquad 0\ otherwise \quad (10)$$

$$M_{\mathrm{pad}}(\mathrm{i}, j) = 1 \text{ if } \mathrm{x_i} = 0 \text{ or } \mathrm{x_j} = 0, \qquad 0\ otherwise \quad (11)$$

$$M_{\mathrm{combined}} = M_{causal} \vee M_{\mathrm{pad}} \quad (12)$$

Token representations begin with learned embeddings followed by layer normalization and dropout regularization, producing initial hidden states $H_0 \in R^{B\times s\times d}$.

**Rotary Positional Encoding for Relative Attention.** To encode positional information without additional parameters, we apply rotary positional embeddings to query and key vectors within each attention head. For head dimension $d_h$ (constrained to be even), we partition each vector into even and odd components and apply position-dependent complex rotations. Given query $q_t$ and key $k_t$ at position $t$ with frequency-based rotation angles $\theta_{t,i} = t/\big(10000^{2i/d_h}\big)$:

$$cos_{t,i} = \cos\big(\theta_{t,i}\big), \qquad sin_{t,i} = \sin\big(\theta_{t,i}\big) \quad (13)$$

The rotational transformation operates on paired dimensions:

$$\tilde{q}_{t,even} = q_{t,even} \odot \cos_t - q_{t,odd} \odot \sin_t \quad (14)$$

$$\tilde{q}_{t,odd} = q_{t,even} \odot \sin_t - q_{t,odd} \odot \cos_t \quad (15)$$

with identical rotations applied to key vectors $\widetilde{k_t}$. This complex-plane rotation imbues dot-product attention with relative positional sensitivity, enabling the model to capture token offset relationships naturally without absolute position dependence—particularly advantageous for short biological sequences where motif positions vary.

**Pre-Normalization Multi-Head Self-Attention.** Each decoder layer employs pre-normalization architecture, applying layer normalization before self-attention computation. For normalized inputs with head dimension $d_k$, attention weights are computed as:

$$A = softmax\big((QK^T)/\sqrt{d_k} + M_{combined}\big) \quad (16)$$

where Q, K, V represent linear projections of layer-normalized inputs, and $M_{combined}$ denotes the composite masking tensor. The attended context $C = A\cdot V$ is merged across attention heads, projected to model dimension $d_{model}$, and combined residually with the original layer input. Pre-normalization stabilizes gradient flow and optimization dynamics, particularly beneficial for mixed-precision training with small batch sizes common in biological sequence modeling.

**Decoder Architecture and Output Projection.** The complete model stacks $L = 8$ identical decoder blocks, each implementing the pre-normalization attention and GEGLU components described above. A final layer normalization produces output representations $H_L \in \mathbb{R}^B \times s \times d$, which are mapped to vocabulary logits through weight-tied projection:

$$logits = H_L \cdot W_E^T \quad (17)$$

where $W_E$ represents the shared token embedding matrix. Weight tying constrains the representational geometry between input and output spaces, improves likelihood calibration, and reduces parameter count—particularly advantageous for modest-scale biological datasets.

**Training Objective and Optimization.** The model optimizes the standard autoregressive language modeling objective with padding-aware masking (Fig. 1d). For target sequence $x$, we minimize masked cross-entropy over valid positions:

$$\mathrm{L} = -\frac{1}{\mathrm{Z}} \cdot \sum_{\mathrm{b,s}} \mathrm{m_{b,s}} \log P_\theta\left(x_{\mathrm{b,s}} \middle| x_{\mathrm{b,<s}}\right) \tag{18}$$

where $m_{b,s} \in \{0,1\}$ masks padding and truncated positions, $Z$ represents the total number of valid tokens, and $x_{b,<s}$ denotes the causal context preceding position $s$ in batch element $b$. We employ AdamW optimization with linear learning rate scheduling including 10% warmup steps. Mixed-precision training and distributed data parallelism via Accelerate ensure computational efficiency and reproducible results.

## Two-Stage Transformer Framework for TCR Full-Length Chain Generation

We developed a hierarchical deep learning framework that decomposes TCR full-length chain generation into two sequential prediction tasks: gene segment identification and complete sequence synthesis. This two-stage approach addresses the inherent complexity of TCR generation for both α and β chains by first predicting variable (V) and joining (J) gene segments from CDR3 sequences, then leveraging this genetic context to generate complete chain sequences. The modular design enables independent optimization while maintaining biological consistency across the generation pipeline for both TCR α and β chains (Fig. 1e).

**Stage 1: Gene Prediction Architecture.** The gene prediction model employs a Transformer encoder to classify V and J gene segments from CDR3 input sequences for both α and β chains. For a tokenized CDR3 sequence $\mathrm{x} = (x_1, \cdots x_s)$ of length $L$, the model constructs initial representations by combining learned amino acid embeddings with positional encodings:

$$H_0 = E_{tok}(x) + E_{pos}(1,2, \cdots L) \tag{19}$$

where $E_{tok} : \mathbb{R}^{v} \rightarrow \mathbb{R}^{d}$ maps vocabulary indices to d-dimensional embeddings and $E_{pos}: \mathbb{N} \rightarrow \mathbb{R}^{d}$ provides learned positional information. The embedded sequence undergoes processing through $N_{enc}$ stacked Transformer encoder layers:

$$H_l = TransformerEncoder_l(H_{l-1}), l = 1, \cdots, N_{\mathrm{enc}} \tag{20}$$

Each encoder layer implements multi-head self-attention with GELU activation and residual connections:

$$Attention(Q, k, V) = softmax\left(Qk^T / \sqrt{d_k}\right) \cdot V \tag{21}$$

$$FFN(H) = GELU(HW_1 + b_1)W_2 + b_2 \tag{22}$$

where $W_1 \in \mathbb{R}^{d \times dff}, W_2 \in \mathbb{R}^{dff \times d}$ with expansion ratio $d_{ff} = 4d$. To obtain sequence-level representations, we apply global average pooling across the sequence dimension:

$$h_{pool} = \frac{1}{L} \cdot \sum_{i=1}^{L} H_{N_{enc},i} \tag{23}$$

The pooled representation feeds into separate classification heads for V and J gene prediction, with distinct vocabularies for α and β chains. Take β chain for example:

$$P_{VB} = softmax\left(W_{VB} h_{pool} + b_{VB}\right) \tag{24}$$

$$P_{JB} = softmax\left(W_{JB} h_{pool} + b_{JB}\right) \tag{25}$$

where $\mathrm{W}_{VB} \in \mathbb{R}^{d \times |V_{\mathrm{VB}}|}$, $\mathrm{W}_{\mathrm{J}B} \in \mathbb{R}^{d \times |V_{\mathrm{J}B}|}$, and $|V_{\mathrm{VB}}|$, $|V_{\mathrm{JB}}|$ represent VB and JB vocabulary sizes, respectively. The training objective combines cross-entropy losses with equal weighting:

$$L_{gene} = CE(P_{VB}, y_{VB}) + CE\left(P_{JB}, y_{JB}\right) \tag{26}$$

**Stage 2: Sequence Generation Architecture.** The sequence generation model implements an encoder-decoder Transformer that integrates CDR3 sequences with predicted gene information to synthesize full-length chains. For β chain, Gene information is embedded through a specialized component that concatenates VB and JB embeddings:

$$g_{VB} = E_{VB}(v), \qquad g_{JB} = E_{JB}(j) \tag{27}$$

$$g_{gene} = LayerNorm(Linear[g_{VB}; g_{JB}]) \tag{28}$$

where $E_{VB}: \mathbb{R}|V_{VB}| \rightarrow \mathbb{R}^{\frac{d}{2}}$ and $E_{JB}: \mathbb{R}\left|V_{JB}\right| \rightarrow \mathbb{R}^{\frac{d}{2}}$ are learned gene embeddings, and $[g_{VB}; g_{JB}]$ denotes concatenation. The encoder processes the concatenated representation of gene context and CDR3β sequence:

$$E_{input} = \left[g_{\mathrm{gene}}; \mathrm{H}_{\mathrm{CDR3}\beta}\right] \tag{29}$$

$$E_{\mathrm{output}} = \mathrm{TransformerEncoder}\left(E_{input}\right) \tag{30}$$

where $\mathrm{H_{CDR3}}$ represents the embedded and positionally-encoded CDR3β sequence. The decoder generates full-length sequences autoregressively using causal attention masking. For target sequence $y = (y_1 \cdots y_M)$, the decoder computes hidden states while attending to encoder outputs:

$$D_t = \mathrm{TransformerDecoder}\left(y_{<t}, E_{output,} M_{causal}\right) \tag{31}$$

where $M_{causal}$ ensures that position $t$ can only attend to previous positions $t' < t$. The output projection layer maps decoder states to vocabulary probabilities:

$$p\left(y_t \middle| y_{<t}, x, g_{VB}, g_{JB}\right) = softmax(W_{out} D_t + b_{out}) \tag{32}$$

where $W_{out} \in \mathbb{R}^{d \times |V_{aa}|}$ projects to the amino acid vocabulary. The sequence generation loss minimizes cross-entropy over valid (non-padding) positions:

$$L_{seq} = -\frac{1}{Z} \cdot \sum_{t=1}^{M} mask_t \cdot \log p\left(y_t \middle| y_{<t}, x, g_{VB}, g_{JB}\right) \tag{33}$$

where $Z$ represents the number of valid positions and $mask_t$ excludes padding tokens from the loss computation.

**Chain-Specific Implementation.** The two-stage design decomposes the complex generation task, improving both gene prediction accuracy and sequence biological plausibility. LSMTCR contains approximately 110 million parameters and utilizes mixed-precision training with gradient accumulation for efficient GPU computation. This modular approach enables independent optimization and evaluation of each component while maintaining end-to-end functionality for complete TCR chain prediction (Fig. 1e).

## IV. Discussion

This work advances epitope-conditioned TCR design by integrating diffusion-style epitope encoding, conditional autoregressive CDR3 generation across both chains, and gene-aware full-length assembly within a single, staged framework. Methodologically, three features are central. First, a time-conditioned, diffusion-enhanced Epitope-BERT improves robustness to weak supervision and unseen epitopes. Second, conditional GPT decoders—pretrained on CDR3β and transferred to CDR3α—provide controllable decoding with temperature scheduling that tunes fidelity–diversity trade-offs; cross-modal conditioning aligns epitope and CDR3 in a shared embedding space, reinforced by length perturbations and a noise curriculum. Third, a TCR-Transformer enforces immunogenetic fidelity by predicting V/J usage and assembling full-length α/β consistent with V/J statistics and pairing distributions.

Empirically, LSMTCR shows several desirable properties relative to representative approaches. For CDR3β, it achieves higher predicted binding probabilities across TEP, MIRA and curated data, with slightly lower performance than GRATCR on McPAS but a tighter, high-score-skewed distribution. It more closely reproduces empirical amino-acid positional frequencies and length distributions, whereas diffusion-based baselines tend to undershoot lengths and grafting-based models over-concentrate on a few lengths. Temperature scaling increases motif coverage, uniqueness, novelty, compositional richness, evenness and length realism showing sensitivity at higher temperatures. Under matched conditions, LSMTCR generally attains higher composite diversity than GRATCR. On CDR3α, transfer learning improves predicted binding, length realism and diversity over GRATCR and TCR-epiDiff, while preserving global positional trends.

For full-length assembly, models seeded with either background or de novo CDR3s produce TCRα/β chains whose length distributions match references, with high exact-match rates, low normalized Hamming and Levenshtein distances, high 3-mer set Jaccard similarities, and small Jensen–Shannon divergences of k-mer spectra. AlphaFold-based paired α/β co-modelling yields higher pTM/ipTM than single-chain settings, in line with the expectation that α/β interfaces determine docking geometry.

These findings are bounded by the evaluation setup. Predicted binding relies on discriminative models and structural proxies; training data are biased toward well-observed events and strong responders; and diversity–fidelity trade-offs depend on temperature. Although retrieval-augmented, soft-constrained decoding and shared embeddings aim to mitigate biases and improve generalization, robustness to rare HLAs, weakly immunogenic epitopes or out-of-distribution contexts remains to be comprehensively assessed. Safety considerations such as cross-reactivity and off-target recognition are not resolved by sequence- or structure-level proxies alone.

In conclusion, LSMTCR operationalizes epitope to full-length α/β design by separating target specificity from immunogenetic constraints and fusing them through staged conditioning and assembly. It improves predicted binding on most benchmarks, recovers biologically grounded length and motif grammars, offers temperature-controlled diversity, and extends generation from CDR3 fragments to full-length, gene-contextualized receptors. By producing diverse TCR αβ candidates from epitope input, LSMTCR supports high-throughput screening, iterative refinement and mechanistic investigation, bringing epitope-to-receptor design closer to practical application.

# Supplementary

Table1 Evaluation for TCRβ Generation Models

| Evaluation Models | Generation Models | Datasets | Average Binding Prediction (%) |
|---|---|---|---|
| NetTCR2 | TCR-epiDiff | MIRA | 62.62 |
| | | TEP | 70.44 |
| | | Mcpas | 53.21 |
| | | Our Dataset | 86.93 |
| | GRATCR | MIRA | 99.95 |
| | | TEP | 99.96 |
| | | Mcpas | **97.43** |
| | | Our Dataset | 99.72 |
| | **LSMTCR** | MIRA | **99.98** |
| | | TEP | **99.97** |
| | | Mcpas | 97.11 |
| | | Our Dataset | **99.83** |
| ATM-TCR | TCR-epiDiff | MIRA | 61.04 |
| | | TEP | 68.23 |
| | | Mcpas | 78.63 |
| | | Our Dataset | 59.49 |
| | GRATCR | MIRA | 100 |
| | | TEP | 100 |
| | | Mcpas | 99.99 |
| | | Our Dataset | 100 |
| | **LSMTCR** | MIRA | **100** |
| | | TEP | **100** |
| | | Mcpas | **100** |
| | | Our Dataset | **100** |
| TEPCAM | TCR-epiDiff | MIRA | 91.24 |
| | | TEP | 91.42 |
| | | Mcpas | 84.34 |
| | | Our Dataset | 94.80 |
| | GRATCR | MIRA | 100 |
| | | TEP | 100 |
| | | Mcpas | 99.96 |
| | | Our Dataset | 99.99 |
| | **LSMTCR** | MIRA | **100** |
| | | TEP | **100** |
| | | Mcpas | **99.96** |
| | | Our Dataset | **99.99** |

Table2 Evaluation for TCRα Generation Models

| ***Evaluation Models*** | ***Generation Models*** | ***Average Binding Prediction (%)*** |
|---|---|---|
| *NetTCR2* | *TCR-epiDiff* | *73.22* |
| | *GRATCR* | *99.25* |
| | ***LSMTCR*** | ***99.63*** |
| *ATM-TCR* | *TCR-epiDiff* | *60.26* |
| | *GRATCR* | *100* |
| | ***LSMTCR*** | ***100*** |
| *TEPCAM* | *TCR-epiDiff* | *88.48* |
| | *GRATCR* | *99.99* |
| | ***LSMTCR*** | ***99.99*** |